\documentclass[12pt,a4paper]{article}

\usepackage{dsfont}

\usepackage{amsmath,amssymb}
\usepackage{listings}
\usepackage{srcltx}
\usepackage{verbatim} 
\usepackage{tikz}
\usepackage{graphicx}
\usepackage[all]{xy} 
\usepackage[bookmarks = true, colorlinks=true, linkcolor = black, citecolor = black, menucolor = black, urlcolor = black]{hyperref}

\definecolor{dkblue}{rgb}{0,0.1,0.5}
\definecolor{lightblue}{rgb}{0,0.5,0.5}
\definecolor{dkgreen}{rgb}{0,0.4,0}
\definecolor{dk2green}{rgb}{0.4,0,0}
\definecolor{dkviolet}{rgb}{0.6,0,0.8}
\definecolor{shadethmcolor}{rgb}{0.9, 0.9,1}

\newcommand{\rrdc}{\mbox{\,\(\Rightarrow\hspace{-9pt}\Rightarrow\)\,}}
\newcommand{\Zset}{\mathbb{Z}}
\newcommand{\wC}{\widehat{C}}

\newtheorem{Thm}{Theorem}

\newtheorem{Defn}[Thm]{Definition}

\newtheorem{Alg}[Thm]{Algorithm}
\newcommand{\inp}{{\itshape Input: }}
\newcommand{\outp}{{\itshape Output: }}

\begin{document}

\title{A certified reduction strategy for homological image processing}
\author{Mar\'\i a Poza, C\'esar Dom\'\i nguez, J\'onathan Heras, Julio Rubio}

\maketitle

\begin{abstract}

The \emph{analysis of digital images} using homological procedures is an outstanding topic 
in the area of Computational Algebraic Topology. In this paper, we describe a \emph{certified}
reduction strategy to deal with digital images, but preserving their homological 
properties. We stress both the advantages of our approach (mainly, the formalisation of the mathematics
allowing us to verify the correctness of algorithms) and some limitations
(related to the performance of the running systems inside proof assistants).
The drawbacks are overcome using techniques that provide an integration
of computation and deduction. Our driving application is a problem in bioinformatics,
where the accuracy and reliability of computations are specially requested.
\end{abstract}

\sloppy

\section{Introduction}

Scientific computing is an outstanding tool to assist researchers in
experimental sciences. When applied to biomedical problems, the
accuracy and reliability of the computations are particularly important.
Thus, the possibility of increasing the trust in scientific software by
means of mechanized theorem proving technology becomes an
interesting area of research. 

In this paper, we explore this path to certify image processing procedures. 
In particular, we have chosen the Coq proof assistant~\cite{Coq} to certify
the programs which allow us to analyse images obtained from neuron cultures~\cite{German}.
The techniques that we use to deal with these images are based on Computational Algebraic Topology.
Nowadays, computing in Algebraic Topology has an increasing
importance in applied mathematics~\cite{EdelHarer}; namely, 
in the context of digital image processing (see~\cite{ADFQ03} and the series of conferences
called \emph{Computational Topology in Image Context}).  
The key observation of our approach is that, after a suitable preprocessing,
the solution of a biological problem (namely, the number of synapsis in
a picture of a neuron) can be identified with the computation of a topological
invariant (the rank of a homology group). Then, all our efforts are concentrated
on computing, in a certified manner, such an invariant.

Since the size of real-life biomedical images is too big to handle them 
directly, we propose a {\em reduction strategy}, that allows us to work with
smaller data structures, but preserving all their homological properties. To this aim,
we use the notion of \emph{discrete vector field} \cite{Forman}, following very
closely an algorithm due to Romero and Sergeraert \cite{RS12}.

In order to verify the correctness of these procedures, it is necessary the formalisation of a certain
amount of mathematics. The most significant piece of mathematics
formalised in this paper is the so-called \emph{Basic Perturbation Lemma} (or
BPL, in short). The proof of this theorem has been already implemented in
the Isabelle/HOL proof assistant \cite{ABR08}. The BPL formalisation presented
in this paper is much shorter and compact than that of \cite{ABR08}. There are
two reasons for this improvement of the formal proof. The former is that in
this work we have followed a new and shorter proof of the BPL (due again
to Romero and Sergeraert \cite{Rom12}). The latter is that we
have built our formal proof on the powerful \emph{SSReflect} library of Coq~\cite{SSReflect}
(on the contrary, much of the infrastructure required was defined from
scratch in \cite{ABR08}). 

Apart from the efficiency in the writing of proofs, using SSReflect also
has other consequences. Since SSReflect is designed to deal only
with \emph{finite structures}, the proof of the BPL presented here only
applies over \emph{finitely generated groups} (the proof formalised in
\cite{ABR08} has not this limitation). Furthermore, dealing with
finite structures, and inside the constructive logic of Coq, eases the
executability of the proofs, and thus the generation of certified programs
(the same tasks in Isabelle/HOL pose more difficulties; see \cite{FAC}). 
However, it is worth mention that this limitation does not mean any 
special hindrance in our work, because digital images are always finite
structures.

In order to prove the correctness of the generated programs, we must
establish, and keep, a link among the initial biomedical picture, and
the final smaller data structure where the homological calculations
are carried out. This implies a big amount of processing, and does not
allow us to execute all the steps \emph{inside} Coq (the full path
has been travelled, but only in \emph{toy} examples). Then we
have appealed to a programming language, Haskell \cite{Haskell} in
our case, to integrate computation and deduction.

Haskell appears in two different steps of  our methodology. In the
early stages of development, Haskell prototypes of the algorithms
are systematically tested by using the  QuickCheck tool \cite{Quickcheck}.
This allows us to discharge many small and common errors, 
which could hinder the proving process in Coq. 
In the final computational step, Haskell is used as an \emph{oracle}
for Coq. The most hard parts of the calculation (in our case,
an important bottleneck is computing inverse matrices) are delegated
to Haskell programs; the \emph{results} of these Haskell programs
are then \emph{proved} correct within Coq.

With this hybrid technique, we have got the objective of computing,
in a certified way, the homology of actual biomedical images coming
from neurological experiments.

The rest of this paper is organized as follows. Section~\ref{image}
is devoted to present a running example, coming from the biomedical
context, as a test-case for our development. The formalisation of an
algorithm to build a discrete vector field associated with a matrix,
using Coq and its SSReflect library,  is explained in
Section~\ref{dvf}. This vector field computation will be used in
Section~\ref{BPL} to reduce the chain complex associated with a
digital image. The reduction process is based on an essential lemma
in Algorithmic Homological Algebra called the Basic Perturbation Lemma.
We also include in that section a proof of such a lemma. In
Section~\ref{computation}, we explain how  the certified programs can
be used to
 effectively compute the homology
 of images. The paper ends with a section of conclusions and further work, and the bibliography.
 
The complete source code of our formalisation can be seen at \url{http://www.unirioja.es/cu/cedomin/crship/}.

\section{Certified image processing}\label{image}
The discipline of Algebraic Digital Topology, or more specifically, the computation of homology groups
from digital images is mature enough (see, for instance,~\cite{Ziou02}, one among many good references)
to go one step further and investigate the possibility of \emph{certified computations} (i.e., computation formally
verified by proving its correctness using
an \emph{interactive} proof assistant) in digital topology, as it happens in other areas of computer
mathematics (see~\cite{FCT}).

\begin{figure}
\centering
\begin{tikzpicture}[scale=.5]
\draw (-3,0.3) node {{\scriptsize Biomedical}};
\draw (-3,-0.3) node {{\scriptsize image}};
\draw (2,0.3) node {{\scriptsize Digital}};
\draw (2,-0.3) node {{\scriptsize image}};
\draw (7,0.3) node {{\scriptsize Cubical/Simplicial}};
\draw (7 ,-0.3) node {{\scriptsize Complex}};
\draw (13,0) node {{\scriptsize Matrix}};
\draw (18.4,0) node {{\scriptsize Homology}};
\draw[-latex] (-1.5,0) --  (1,0);
\draw[-latex] (3,0) --  (5.5,0);
\draw[-latex] (8.75,0) --  (11.5,0);
\draw[-latex] (14.5,0) -- (17,0);
\draw[rounded corners=20pt,-latex] (12,-.25) -- (13,-1.5) -- (14,-.25);
\draw[rounded corners=100pt,-latex] (18.5,.5) -- (8,2.5) -- (-3,.5);
\draw (13,-1.25) node {{\scriptsize reduction}};
\draw (8,2.55) node[anchor=north] {{\scriptsize interpretation}};
\end{tikzpicture}
\caption{Computing homology from a digital image.}\label{chfadi}
\end{figure}
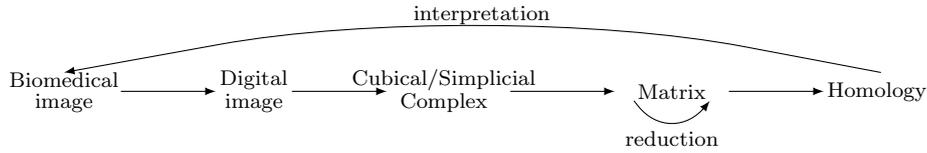

In a very rough manner, the process to be verified is reflected in Figure~\ref{chfadi}.
Putting it into words, we firstly pre-process a biomedical image to obtain a monochromatic image. 
From the black pixels of such a monochromatic image a cubical/simplicial complex is obtained
(by means of a triangulation procedure); subsequently, from the cubical/simplicial complex, its \emph{boundary (or incidence) matrices}
are constructed, and finally, \emph{homology} can be computed. If we work with coefficients over
a field (and it is well-known that it is enough to take as coefficients the field $\mathbb{Z}_2$,
when working with 2D and 3D digital images) and if only the \emph{degrees} of the homology groups (as
vector spaces) are looked for, then having a program
able to compute the rank of a matrix is sufficient to accomplish the whole task. In this process the matrix obtained from the image can be huge.
In this case, a process of reduction of the matrix without loosing the homological properties of the image can be applied previously to compute the homology. 
This architecture is particularised  in this paper with a real problem that appeared in a biomedical application
and with the Coq proof assistant and its SSReflect library as programming and verifying tool.

Biomedical images are a suitable benchmark for testing our programs, the reason is twofold. First of all, the amount of information
included in this kind of images is usually quite big. Then, a process able to reduce those images but keeping
the homological properties can be really useful. Secondly, software systems dealing with biomedical
images must be trustworthy. This is our case since we have formally verified the correctness of our programs.

As an example, we can consider the problem of counting the number of \emph{synapses} in a neuron. Synapses~\cite{Neuroscience}
are the points of connection between neurons and are related to the computational capabilities of the brain.  The possibility of changing the number of synapses may be an important asset
in the treatment of neurological diseases, such as Alzheimer, see~\cite{Alzheimer}. Therefore,
we can claim that an efficient, reliable and automatic method to count synapses is
instrumental in the study of the evolution of synapses in scientific experiments.

\begin{figure}
\centering
 \includegraphics[scale=0.14]{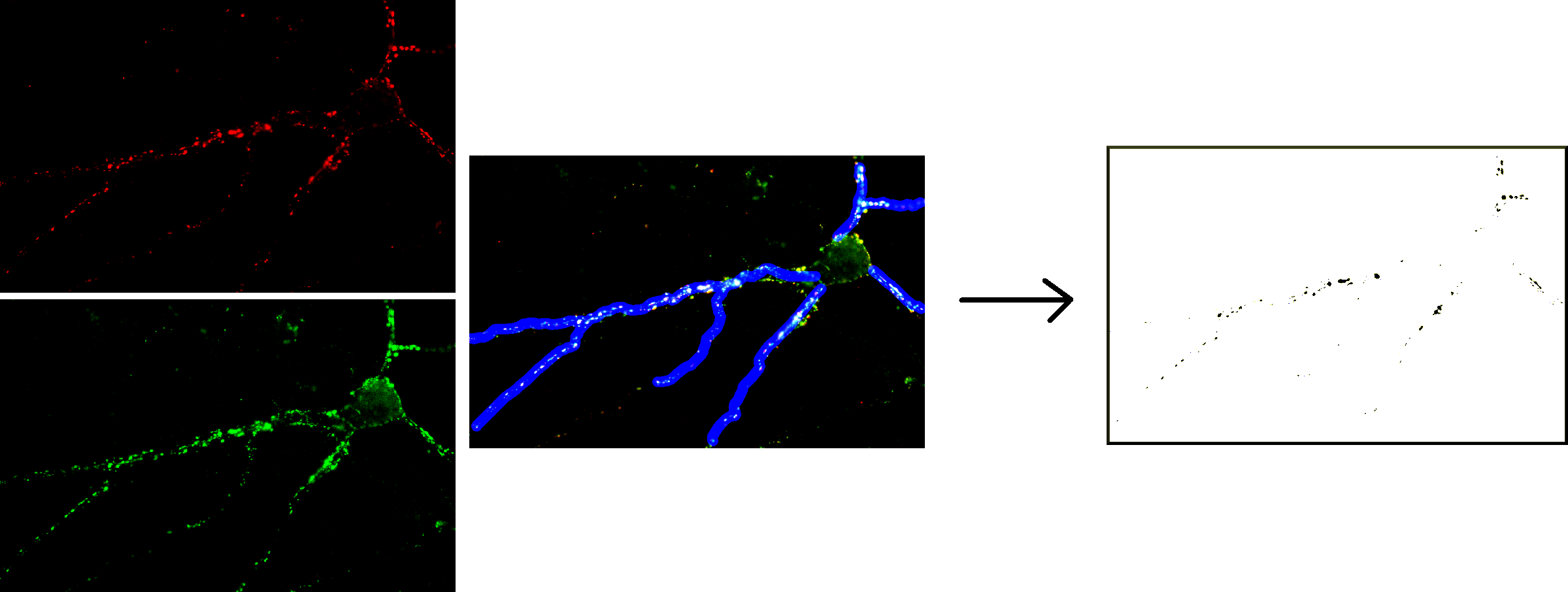}
\caption{Synapses extraction from three images of a neuron.}\label{fig:neuron}
\end{figure}

Up to now, the study of the \emph{synaptic density evolution} of
neurons was a time-consuming task since it was performed, mainly,
manually. To overcome this issue, an automatic method was presented
in~\cite{HMPR11-ACA}. Briefly speaking, such a process can be split
into two parts. Firstly, from three images of a neuron (the neuron
with two antibody markers and the structure of the neuron), we obtain a
monochromatic image, see
Figure~\ref{fig:neuron}\footnote{The same images with higher
resolution can be seen in
\texttt{http://www.unirioja.es/cu/joheras/synapses/}}. In such an
image, each connected component represents a synapse. So, the
problem of measuring the number of synapses is translated into a
question of counting the connected components of a monochromatic
image.

In the context of Algebraic Digital Topology, this issue can be
tackled by means of the computation of the homology group $H_0$ of
the monochromatic image. This task can be performed in Coq
through the formally verified programs which were presented
in~\cite{HDMMPS}. Nevertheless, such programs are not able to handle
images like the one of the right side of Figure~\ref{fig:neuron} due
to its size. It is worth noting that Coq is a Proof Assistant
and not a Computer Algebra system; and, in general, efficiency of algorithms is pushed into the
background of this kind of systems. However, there is an effort towards the efficient implementations of
mathematical algorithms running inside Coq, as shown by recent works on efficient real numbers
\cite{krebbers_computer_2011}, machine integers and arrays
\cite{armand_extending_2010} or an approach to compiled
execution of internal computations \cite{gregoire_compiled_2002}.


In our case, we apply a reduction process of the data structures but
without loosing the homological properties to overcome the efficiency problem.
In particular, we have
focussed on the formalisation of \emph{discrete vector fields}, a
powerful notion that has been welcomed in the study of homological
properties of digital images, see~\cite{Cazals2003,GBHP08,Jerse2010}.  The importance of
discrete vector fields, which were first introduced
in~\cite{Forman}, stems from the fact that they can be used to
considerably reduce the amount of information of a discrete object
but preserving homological properties. Using this approach, we can
successfully compute the homology of the previous biomedical image
in just $25$ seconds, a remarkable time for an execution inside {\sc
Coq}. Besides, we have proved using this proof assistant that the
homological properties of the initial digital image and the reduced
one are preserved.

In this section, we include the basic definitions which,
mainly, come from the algebraic setting of Discrete Morse Theory
presented in~\cite{RS12}. In addition, we present an algorithm to
construct an admissible discrete vector field from a matrix. Then, a
formalisation in Coq of this algorithm is provided. Finally,
we introduce the fundamental notions in the Effective Homology
theory~\cite{RS02} and state the theorem where Discrete Morse
Theory and Effective Homology converge.

\subsection{Basic mathematical definitions}
We assume as known the notions of \emph{ring}, \emph{module} over a ring and \emph{module morphism} (see, for instance,~\cite{Basic-Algebra}).
First of all, let us introduce one of the main notions in the context of Algebraic Topology: \emph{chain complexes}.

\begin{Defn}
A \emph{chain complex} $C_\ast$ is a pair $(C,d)$, where
$C=\{C_n\}_{n\in\mathbb{Z}}$ is a family of ${\mathcal R}$-modules
and $d=\{d_n:C_n\rightarrow C_{n-1}\}_{n\in\mathbb{Z}}$ is family of
module morphisms, called the \emph{differential map}, such that
$d_{n-1}\circ d_n= 0$, for all $n\in\mathbb{Z}$. In many situations
the ring ${\mathcal R}$ is either the integer ring, ${\mathcal R}
=\mathbb{Z}$, or the field $\mathbb{Z}_2$. Usually, we denote the
chain complex $C_\ast=(C_n,d_n)_{n\in\mathbb{Z}}$. A chain complex
is \emph{free} (\emph{finitely generated}) if its modules are free
(finitely generated).
\end{Defn}

The image
$B_n = im~d_{n+1} \subseteq C_n$ is the (sub)module of \emph{$n$-boundaries}. The kernel $Z_n = \ker d_n \subseteq C_n$ is the
(sub)module of \emph{$n$-cycles}.
Given a chain complex $C_\ast=(C_n,d_n)_{n \in \mathbb{Z}}$, the identities $d_{n-1}\circ d_n= 0$ mean the inclusion
relations \(B_n \subseteq Z_n\): every boundary is a cycle (the converse in general is not true). Thus the next definition makes sense.

\begin{Defn}
The \emph{$n$-homology group} of $C_\ast$, denoted by $H_n(C_\ast)$, is defined as the quotient $H_n(C_\ast)=Z_n/B_n$.
\end{Defn}

Chain complexes have a corresponding notion of morphism.

\begin{Defn}
Let $C_\ast=(C_n,d_n)_{n\in\mathbb{Z}}$ and
$D_\ast=(D_n,\widehat{d}_n)_{n\in\mathbb{Z}}$ be two chain
complexes. A \emph{chain complex morphism} $f:C_\ast \rightarrow
D_\ast$ is a family of module morphisms, $f=\{f_n:C_n \rightarrow
D_n\}_{n\in\mathbb{Z}}$, satisfying  the relation $f_{n-1}\circ d_n
= \widehat{d}_n\circ f_n$, for all $n\in\mathbb{Z}$. Usually, the
sub-indexes are skipped, and we just write $f\circ d =
\widehat{d}\circ f$.
\end{Defn}

Let us state now the main notions coming from the algebraic setting
of Discrete Morse Theory~\cite{RS12}.

\begin{Defn}
Let $C_\ast=(C_n,d_n)_{n \in \Zset}$ be a free chain complex with a distinguished $\Zset$-basis $\beta_n \subset C_n$, for all $n\in\mathbb{Z}$.
A \emph {discrete vector field} $V$ on $C_\ast$ is a collection of pairs $V = \{(\sigma_i, \tau_i)\}_{i\in I}$ satisfying the conditions:
\begin{itemize}
\item Every $\sigma_i$ is some element of $\beta_n$, in which case $\tau_i\in \beta_{n+1}$.
The degree $n$ depends on $i$ and in general is not constant.
\item Every component $\sigma_i$ is a \emph{regular face} of the corresponding $\tau_i$ (regular face means that the coefficient of
$\sigma_i$ in $d_{n+1}(\tau_i)$ is $1$ or $-1$).
\item Each generator (\emph{cell}) of $C_\ast$ appears at most one time in $V$.
\end{itemize}
\end{Defn}

It is not compulsory that all the cells of $C_\ast$ appear in the vector field $V$.

\begin{Defn}
A cell $\chi$ which does not appear in a discrete vector field $V=\{(\sigma_i, \tau_i)\}_{i\in I}$ is called a \emph{critical cell}. The elements $\sigma_i$ and $\tau_i$ in the vector field are called \emph{source} and \emph{target cells}, respectively.
\end{Defn}

From a discrete vector field $V$ on a chain complex, we can
introduce the notion of $V$-paths.

\begin{Defn}
A $V$-path of degree $n$ and length $m$ is a sequence $((\sigma_{i_k}, \tau_{i_k}))_{0 \leq k < m}$ satisfying:
\begin{itemize}
\item Every pair $(\sigma_{i_k}, \tau_{i_k})$ is a component of $V$ and $\tau_{i_k}$ is an $n$-cell.
\item For every $0 < k < m$, the component $\sigma_{i_k}$ is a face of $\tau_{i_{k-1}}$ (the coefficient of $\sigma_{i_k}$ in $d_{n}(\tau_{i_{k-1}})$ is non-null)
different from $\sigma_{i_{k-1}}$.
\end{itemize}
\end{Defn}

\begin{Defn}\label{def:admis}
A discrete vector field $V$ is \emph{admissible} if for every $n \in \Zset$, a function $\lambda_n: \beta_n \rightarrow \mathbb{N}$ is
provided satisfying the following property: every $V$-path starting from $\sigma \in \beta_n$ has a length bounded by $\lambda_n(\sigma)$.
\end{Defn}

In this way, infinite paths are avoided in an admissible discrete vector field.
We will see that when a discrete vector field is admissible it is possible to distinguish the ``useless'' elements of the chain complex (in the sense, that they can be removed without
changing its homology) and the \emph{critical} ones (those whose removal could modify the homology).

If we consider the case of \emph{finite type} chain complexes, where
there is a finite number of generators in each degree, the
differential maps can be represented as matrices. In that case, the
problem of finding an admissible discrete vector field on the chain
complex can be solved through the computation of an admissible
vector field for those matrices.

\begin{Defn}\label{def:dvfM}
Let $M$ be a matrix with coefficients in $\Zset$, and with $m$ rows
and $n$ columns. A \emph{discrete vector field} $V$ for $M$
is a set of natural pairs $\{(a_i, b_i)\}_{i\in I}$ satisfying, for
all $i\in I$, the following conditions:
\begin{itemize}
 \item $1\leq a_i \leq m$ and $1\leq b_i \leq n$.
 \item $M[a_i,b_i]= \pm 1$.
 \item The indexes $a_i$ (resp. $b_i$) are pairwise different.
\end{itemize}
\end{Defn}

When we work in a finite context the admissibility property is
obtained if we avoid loops in the paths. Let $V$ be a vector field
for a matrix $M$ with $m$ rows and $n$ columns.  If $1\leq a,a'\leq
m$, with $a\neq a'$ are two source cells, we can decide $a > a'$ if
there is an elementary $V$-path from $a$ to $a'$, that is, if a
vector $(a,b)$ is present in $V$ and $M[a',b]$ is non-null. This
means that $a$ is a regular face and $a'$ is an arbitrary face of
$b$. In this way, a binary relation is obtained on source cells.
Then, the vector field $V$ is \emph{admissible} if and only if this
binary relation transitively generates a partial order, that is, if
there is no loop $a_1 > a_2 > \ldots > a_k = a_1$.

\subsection{Romero-Sergeraert's algorithm}
\label{sec:algorithm:rsa}

All the algorithms devoted to construct admissible discrete vector
fields share the same goal: the construction of an admissible
discrete vector field as big as possible (see for instance
\cite{Kozlov07} or \cite{LLT04}). Some of them return a vector field
quickly, but others spend more time to compute it. Let us emphasize
that the latter ones are notable because the search has been more
thorough, so the number of vectors will be higher. We are interested in an algorithm which not only gives us a big vector field but also
does not spend too much time to compute it. The Romero-Sergeraert algorithm (from now on RS algorithm)
does not always build the biggest vector field possible, but the number
of vectors is quite close to the biggest one. Furthermore, in many
cases, it returns the best possible vector field. Moreover,
it is fast enough to obtain the vector field in our application
domain. Due to these reasons, the RS algorithm has been chosen to make our computations.

Briefly, the RS algorithm builds an admissible discrete vector field running the rows of a matrix. It looks for the first element in the
row which verifies the admissibility property with respect to the previously elements included in the discrete vector filed. We define the RS algorithm as follows.

\begin{Alg}[The RS Algorithm]\label{alg:rs}
\textcolor{white}{.}\\
\inp a matrix $M$ with coefficients in $\mathbb{Z}$.\\
\outp an admissible discrete vector field $V$ for $M$ and a list of relations $r$.\\
{\itshape Description: }
  \begin{enumerate}
    \item Initialise the vector field $V$ to the void vector field and the relations $r$ to empty.
    \item For every row $i$ of $M$:
      \begin{enumerate}
        \item For every column $j$, which is different from the second components of $V$, such that $M[i,j]=1$ or $M[i,j]=-1$:
        \begin{itemize}
        \item[] Look for the rows $k\neq i$ such as $M[k,j]\neq 0$ and obtain the relations $i>k$. Then, build the transitive closure of $r$ and these relations.
            \begin{itemize}
                \item[]\verb"If" there is no loop in that transitive closure:
                \begin{itemize}
                    \item[] \verb"then": Add $(i,j)$ to $V$, let $r$ be that transitive closure, and repeat from Step 2 with the next row.
                    \item[] \verb"else": Repeat from Step 2.1. with the next column.
                \end{itemize}
              \end{itemize}
         \end{itemize}
      \end{enumerate}
  \end{enumerate}
\end{Alg}

In general, this algorithm can be applied over matrices with
coefficients in a ring. In that case, the possible vectors will be
only the elements whose value is a unit of the ring. Specifically,
if we work with a field $F$, instead of a ring, every non-null
element is a unit. In our particular case,
as the homology groups of 2D images are torsion-free, we will work with the
field $\mathbb{Z}_2$. Therefore, the selected
vectors are entries whose value is $1$. From now on, we will work
with $\mathbb{Z}_2$.

Let us mention that it is extremely relevant sorting the admissible
discrete vector field because we will sort the matrix as a previous
step to reduce it. For every vector $(a,b)$, the
value of the function $\lambda(a)$, which gives us the longest path
from $a$, is computed. In our case, as we build the transitive closure, it is the
maximum length of the relations which start with $a$. Then, we sort
the vector field by the values of $\lambda$ in decreasing order.
Then, the rows and columns of the matrix are sorted using the
ordered vector field. This matrix will have a lower
triangular matrix with 1's in the diagonal as upper-left submatrix
of dimension the number of vectors in the vector field.

\subsection{Formalisation of the RS algorithm in SSReflect}\label{subsec:frss}


The development of a formally certified implementation of the RS
algorithm was presented in \cite{Calculemus2012}. It followed the
methodology presented in~\cite{Mortberg2010}. Firstly, the programs
were implemented in Haskell~\cite{Haskell}, a \emph{lazy}
functional programming language. Subsequently, our implementation
was intensively tested using QuickCheck~\cite{Quickcheck}, a
tool which allows one to automatically test properties about
programs implemented in Haskell. Finally, the correctness of
our programs were verified using the Coq~\cite{Coq} proof
assistant and its SSReflect library~\cite{SSReflect}. In this
section, we briefly show the last step of this process.

First of all, we define the data types related to our programs. A matrix is represented by means of a list of lists of the same length over $\mathbb{Z}_2$
(encoded using the unit ring type associated with the booleans), a vector field is a sequence of natural pairs, and finally, the relations are a list of 
lists of natural numbers. The notation \lstinline?a:T? means that the variable \lstinline?a? has type \lstinline?T?.

\begin{lstlisting}
  Definition Z2:= bool_cunitRingType.
  Record matZ2:=
    {M: seq (seq Z2);
     m: nat;
     n: nat;
     is_matrix:  M = [::] \/
       [/\ m = size M & forall i, i < m -> size (rowseqmx M i) = n]}.
  Definition vectorfield:= seq (prod nat nat).
  Definition rels:= seq(seq nat).
\end{lstlisting}

Then, we have defined a function called \lstinline?Vecfieldadm? that checks
if an element \lstinline?vf: vectorfield? satisfies the properties of
an admissible discrete vector field with respect to a matrix
\lstinline?M: matZ2? and a list of relations \lstinline?r: rels?.
\begin{lstlisting}
  Definition Vecfieldadm (M: matZ2)(vf: vectorfield)(r:rels):=
    (all [pred i | 0<= i < (m M)](getfirstseq vf)) /\
    (all [pred i | 0<= i < (n M)](getsndseq vf)) /\
    (forall i j:nat, (i,j) \in vf -> (nth 0 (nth nil M i) j) = 1%R) /\
    (uniq (getfirstseq vf)) /\ 
    (uniq (getsndseq vf)) /\
    (forall i j l:nat, (i,j) \in vf -> i!=l
       -> (nth 0 (nth nil M l) j)!= 0%R -> (i::l::nil) \in r) /\
    (forall a b s p, (a::s) \in r -> (last 0%N s = b)
       -> (b::p) \in r -> ((a::s) ++ p) \in r) /\
    (all uniq r) /\ 
    (ordered glMax vf r).
\end{lstlisting}

Let us note that the first five conditions come from the three properties of a discrete vector field (see Definition \ref{def:dvfM}).
The next three conditions are linked with the relations. The first one gives us the link between the vector field and the relations.  
The second one verifies that we are constructing the transitive closure. And the last one states the admissibility property.
It makes sure that every sequence of \lstinline?r? has not repeated elements. Finally, it is checked that \lstinline?vf? is ordered taking
into account the \lstinline?glMax? function, which returns the longest path associated with a cell, and the relations \lstinline?r?.

The RS algorithm has been implemented using two functions: \lstinline?genDvf?, which constructs an admissible discrete
vector field, and \lstinline?genOrders?, which generates the relations. As we have explained in the last paragraph of the previous
subsection, the discrete vector field generated by the RS algorithms must be reordered, this is achieved thanks to the function 
\lstinline?dvford?. The correctness of this program has been proved in the theorem \lstinline?dvfordisVecfieldadm? -- this theorem 
establishes that given a matrix \lstinline?M?, the output produced by \lstinline?dvford? satisfies the properties specified in
\lstinline?Vecfieldadm?.

\begin{lstlisting}
 Theorem dvfordisVecfieldadm (M:matZ2):
    Vecfieldadm M (dvford M)(genOrders M).
\end{lstlisting}

The proof of the above theorem has been split into a series of lemmas
which correspond with each one of the properties that should be
fulfilled to have an admissible discrete vector field. For instance,
the lemma associated with the first property of the definition of a
discrete vector field is the following one.

\begin{lstlisting}
  Lemma propSizef (M:matZ2):
   (all [pred i | 0 <= i < (size M)](getfirstseq (genDvf M)))
\end{lstlisting}

Both the functions which implement the RS algorithm and the ones
needed to specify the properties of admissible discrete
vector fields are defined using a \emph{functional style}; that is,
our programs are defined using \emph{pattern-matching} and
\emph{recursion}. Therefore, in order to reason about our recursive
functions, we need elimination principles which are fitted for them.
To this aim, we use the tool presented in~\cite{BC02} which allows
one to reason about complex recursive definitions in Coq.

For instance, in our development of the implementation of the RS
algorithm, we have defined a function, called \lstinline?subm?,
which takes as arguments a natural number \lstinline?n? and a matrix
\lstinline?M? and removes the first \lstinline?n? rows of
\lstinline?M?. The inductive scheme associated with \lstinline?subm?
is set as follows.

\begin{lstlisting}
  Functional Scheme subm_ind := Induction for subm Sort Prop.
\end{lstlisting}

Then, when we need to reason about \lstinline?subm?, we can apply
this scheme with the corresponding parameters using the instruction
\lstinline?functional induction?. However, as we have previously
said both our programs to define the RS algorithm and the ones which
specify the properties to prove are recursive. Then, in several
cases, it is necessary to merge several inductive schemes to
induction simultaneously on several variables.  For instance, let
$M$ be a matrix and $M'$ be a submatrix of $M$ where we have removed
the $(k-1)$ first rows of $M$; then, we want to prove that $\forall
j,\; M (i, j) = M' (i - k + 1, j)$. This can be stated in Coq
as follows.

\begin{lstlisting}
  Lemma Mij_subM (i k: nat) (M: matZ2):
    k <= i -> k != 0 -> let M' := (subm k M) in
    M i j ==  M' (i - k + 1) j.
\end{lstlisting}

To prove this lemma it is necessary to induct simultaneously on the
parameters \lstinline?i?, \lstinline?k?, and \lstinline?M?, but the
inductive scheme generated from \lstinline?subm? only applies
induction on \lstinline?k? and \lstinline?M?. Therefore, we have to
define a new recursive function, called \lstinline?Mij_subM_rec?, to
provide a proper inductive scheme.

\begin{lstlisting}
  Fixpoint Mij_subM_rec (i k: nat) (M: matZ2) :=
  match k with
  |0 => M
  |S p => match M with
        |nil => nil
        |hM::tM => if (k == 1)
                    then a::b
                    else (Mij_subM_rec p (i- 1) tM)
        end
  end.
\end{lstlisting}

This style of proving functional programs in Coq is the one
followed in the development of the proof of Theorem
\lstinline?dvfordisVecfieldadm?.


\subsection{Vector Field Reduction theorem}

In this section, we state the theorem where Discrete Morse Theory and Effective Homology converge. First, we introduce \emph{reductions}, one of
the fundamental notions in the Effective Homology theory.

\begin{Defn}
\label{defn:reduction} A \emph{reduction} $\rho$ between two chain
complexes $C_\ast=(C_n,d_n)_{n \in \Zset}$ and $D_\ast=(D_n,\hat{d}_n)_{n \in \Zset}$, denoted by
$\rho: C_\ast \rrdc D_\ast$, is a triple $\rho=(f,g,h)$ where
$f:C_\ast\rightarrow D_\ast$ and $g:D_\ast\rightarrow C_\ast$ are
chain complex morphisms, $h = \{h_n:C_n \rightarrow
C_{n+1}\}_{n\in\mathbb{Z}}$ is a family of module morphism, and the
following properties are satisfied:
\begin{enumerate}
\item [1)]$f \circ g = id$;
\item [2)]$g\circ f + d\circ h + h\circ  d = id$;
\item [3)]$f \circ h = 0$; \quad $h \circ g = 0$; \quad $h \circ h = 0$.
\end{enumerate}
\end{Defn}

The importance of reductions lies in the fact that given a reduction
$\rho~:~C_\ast \rrdc D_\ast$, then $H_n(C_\ast)$ is isomorphic to
$H_n(D_\ast)$ for every $n\in \mathbb{Z}$. Very frequently, $D_\ast$
is a much smaller chain complex than $C_\ast$, so we can compute the
homology groups of $C_\ast$ much faster by means of those of
$D_\ast$.

\begin{Thm}[Vector Field Reduction Theorem]\label{thm:advred}
Let $C_\ast=(C_n,d_n)_{n \in \Zset}$ be a free chain complex and $V$ be an admissible discrete vector
field on $C_\ast$. Then, the vector field $V$ defines a canonical reduction $\rho : (C_n, d_n)_{n \in \Zset}\rrdc (C^c_n, d^c_n)_{n \in \Zset}$ where $C^c_n = \Zset[\beta^c_n]$
is the free $\Zset$-module generated by $\beta^c_n$, the critical $n$-cells of $C_\ast$.
\end{Thm}

\noindent Therefore, the bigger the admissible discrete vector field $V$ the smaller the chain complex $C^c_\ast$.

A quite direct proof of the Vector Field Reduction Theorem based on
the Basic Perturbation Lemma appeared in \cite{RS12}.

\begin{Thm}[Basic Perturbation Lemma, BPL]
Let $\rho=(f,g,h): (C, d) \rrdc (\widehat{C},\widehat{d})$ be a
reduction, and $\delta$ be a \emph{perturbation} of $d$, that is,
$\delta$ is a morphism such that $d+\delta$ is a differential map
for $C$. Furthermore, the function $\delta\circ h $ is assumed
\emph{locally nilpotent}, in other words, for every $x \in C$ there
exists $m \in \mathbb{N}$ (in general, $m$ can depend on $x$) 
such that $(\delta\circ h)^m(x)=0$. Then,
a perturbation $\widehat{\delta}$ can be defined for the
differential $\widehat{d}$ and a new reduction $\rho': (C, d
+\delta) \rrdc (\widehat{C}, \widehat{d} + \widehat{\delta}) $ can
be constructed.
\end{Thm}

The proof of the Vector Field Reduction theorem based on the BPL
is quite simple. Let us explain this proof in a nutshell. First of all,
we consider a particular chain complex $(C,\delta)$ with the same underlying
graded module than $(C, d)$, but with a simplified differential map. Namely,
each component $\delta_n: C_n \rightarrow C_{n-1}$ is defined in the
following way. It is clear that the vector field $V$ defines a
canonical decomposition of $C_n$ depending on the generators are
source, target, or critical cells. Then, $\delta_n$ apply to $0$ the
source and critical cells. If $\tau$ is a target cell, there is a
unique vector $(\sigma, \tau)\in V$. Then, $\delta_n(\tau)=
\varepsilon(\sigma, \tau) \sigma$, where $\varepsilon(\sigma, \tau)$
is the coefficient $1$ or $-1$ of $\sigma$ in $d_n(\tau)$. A
homotopy operator $h$ is defined in the same way as $\delta$ but in
the reverse direction.

Then, we can define an initial reduction,
$\rho=(f,g,h):(C,\delta)\rrdc (C^c, 0)$, of the previous chain
complex to the chain complex  which modules are generated by the
critical cells and which differential map is null. Now, let us
define the morphism $P = (d -\delta)$ which is clearly a
perturbation of $\delta$.  If the nilpotency condition is satisfied,
then the BPL produces the required reduction. The composition $(d
-\delta)\circ h$ is not null only for source cells. For a source
cell $\sigma$, the images  $((d -\delta)\circ h)^m (\sigma)$ 
correspond to walking $V$-paths starting at this cell. As the
vector field is admissible, the length of these paths is bounded,
the image goes eventually to zero, and the nilpotency condition is obtained.

Again, in the case of finite type chain complexes which
differentials are represented as matrices, given an admissible
discrete vector field for those matrices, we can construct new
matrices taking into account the critical components. These smaller
matrices define chain complexes which preserve the homological
properties. This is the equivalent version of Theorem~\ref{thm:advred} for
finite type chain complexes. A detailed description of the process
can be seen in~\cite{RS12}.

\section{Basic Perturbation Lemma}\label{BPL}

In this section, we introduce a formalisation in SSReflect of
the BPL, a central lemma in Algorithmic Homological Algebra 
-- in particular, it has been intensively used in the Kenzo Computer Algebra system~\cite{Kenzo}.
In the literature, there are
several ways of proving this lemma (see, for instance
\cite{BarnesLambe91, RS97}). There are also works related to the
formalisation of the BPL. For instance, the non-graded case of this
lemma is proved in \textsl{Isabelle/HOL} \cite{ABR08}. Furthermore,
a particular case of the BPL is also proved in \textsl{Coq} using
bicomplexes \cite{DR2011}. Now, we show a formalisation of the general
case in SSReflect but with finitely generated structures. Let
us recall that SSReflect only works with finite types; so, it can
seem that this technology can restrict our development. Nevertheless,
this approach is enough because we are interested in applying the BPL to the
computation of the homology associated with a digital image -- a case where every chain
group is finitely generated. Indeed, in this context, this lemma is
applied to a reduction where most of the differentials are null.


\subsection{Mathematical proof of the BPL}
The proof of the BPL (explained in \cite{Rom12}) is based on two results. The former, named 
Decomposition Theorem, builds a decomposition of a chain complex from a reduction of it. The latter 
is a Generalisation of the Hexagonal Lemma.

\begin{Thm}[Decomposition Theorem]
\label{prop:dt} Let $\rho = (f,g,h): (C,d) \rrdc (\wC, \hat{d})$ be
a reduction. This reduction can be used to obtain a decomposition $C = A
\oplus B \oplus C'$ where: $C' = im\; g$, $A \oplus B = \ker f$, $A
= \ker f \cap \ker h$, $B = \ker f \cap \ker d$, the chain complex
morphisms $f$ and $g$ are inverse isomorphisms between $C'$ and
$\wC$, and the arrows $d$ and $h$ are module isomorphisms between
$A$ and $B$.
\end{Thm}

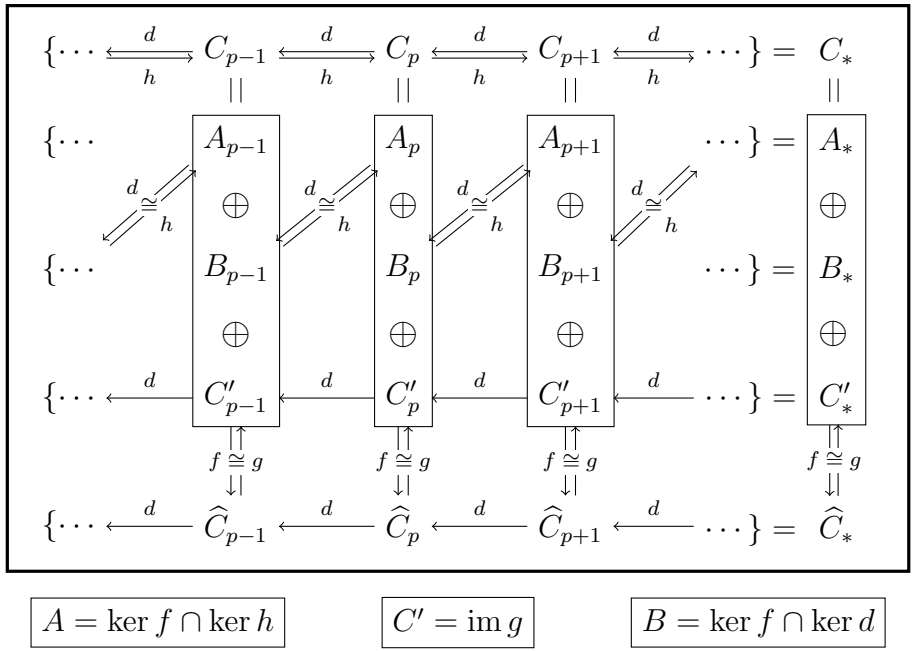
\begin{figure}[h]
\begin{equation}
  \begin{tikzpicture} [xscale = 2.2, yscale = 1.7, baseline = (n22)]
  \foreach \i/\ic in {0,1,2,3,4/4.25}
    {\foreach \j/\jc in {0,1,2,3,4/3.7} {\coordinate (\i\j) at (\ic,\jc) ;}}
  \foreach \j in {0,1,2,3,4} {\node (n0\j) at (0\j)
     {\(\{\cdots\)} ;}
  \foreach \i/\it in {1/p-1, 2/p, 3/p+1}
    {\foreach \j/\jt in {0/\wC, 1/C', 2/B, 3/A, 4/C}
      {\node (n\i\j) at (\i\j) {\(\jt_{\it}\)} ;}}
  \foreach \j/\jt in {0/\wC, 1/C', 2/B, 3/A, 4/C}
    {\node (n4\j) at (4\j) {\(\cdots\} =
    \hspace{5pt} \jt_\ast\)} ;}
  \begin{scope} [font = \scriptsize]
    \path (n14.base) coordinate (b) +(0,0.05) coordinate (bb) ;
    \foreach \i/\ii in {0/1, 1/2, 2/3, 3/4}
      {\draw [->] (b -| n\i4.east) -- node [below] {\(h\)} (b -| n\ii4.west) ;
       \draw [<-] (bb -| n\i4.east) -- node [above] {\(d\)} (bb -| n\ii4.west)
       ;}
    \foreach \i/\ii in {02/13, 12/23, 22/33, 32/43}
      {
      \path (n\i.north east) +(135:0.03) coordinate (1l) +(-45:0.03) coordinate (1r) ;
      \path (n\ii.south west) +(135:0.03) coordinate (2l) +(-45:0.03) coordinate (2r) ;
      \draw [->] (2l) -- (1l) ; \draw [->] (1r) -- (2r) ;
      \path (n\i.north east) --
      node [fill = white, inner sep = 1, label=135:\(d\),
      label = -45:\(h\)] {\(\cong\)} (n\ii.south west) ;
      }
    \foreach \i in {1,2,3}
      {
       \draw (n\i1.south west) rectangle (n\i3.north east) ;
       \path (\i1) -- node {\large\(\oplus\)} (\i2) -- node {\large\(\oplus\)} (\i3) ;
      }
    \foreach \i/\j in {1/2, 2/3}
      {
       \path (4\i) -- coordinate (c) (4\j) (c) +(0:0.32) node {\large\(\oplus\)};
      }
    \path (n43.north east) +(180:0.35) coordinate (c) ;
    \draw (n43.north east) rectangle (c |- n41.south) ;
    \foreach \i/\ii in {1/0,2/1,3/2,4/3} \foreach \j in {0,1}
      {\draw [->] (n\i\j) -- node [above] {\(d\)} (n\ii\j) ; }
    \foreach \i in {1,2,3}
      {
       \path (n\i0.north) +(180:0.03) coordinate (c1) +(0:0.03) coordinate (c2) ;
       \path (n\i1.south) +(180:0.03) coordinate (c3) +(0:0.03) coordinate (c4) ;
       \draw [->] (c2) -- (c4) ; \draw [->] (c3) -- (c1) ;
       \path (\i0) -- node [fill = white, inner sep = 1] {\(f \cong g\)} (\i1) ;
       \draw (c1 |- n\i3.north) +(90:0.1) -- (c1 |- n\i4.south)
             (c2 |- n\i3.north) +(90:0.1) -- (c2 |- n\i4.south) ;
      }
    \path (n40.north) ++(0:0.3) coordinate (c1) +(0:0.05) coordinate (c2) ;
    \draw [->] (c2) -- (c2 |- n41.south) ; \draw [->] (c1 |- n41.south) -- (c1) ;
    \path (c1) -- node [fill = white, inner sep = 1] {\(f \cong g\)} (c2 |- n41.south) ;
    \draw (c1 |- n43.north) +(90:0.1) -- (c1 |- n44.south)
          (c2 |- n43.north) +(90:0.1) -- (c2 |- n44.south) ;
\end{scope}
  \path (n00.south west) +(-135:0.2) coordinate (c1)
        (n44.north east) +(30:0.3) coordinate (c2) ;
  \draw [very thick] (c1) rectangle (c2) ;
  \path (c1) +(-90:0.4) coordinate (c1) ;
  \path (c1) -- node [draw, pos=0.167] {\(A = \ker f \cap \ker h\)}
                node [draw, pos = 0.5] {\(C' = \textrm{im}\,g\)}
                node [draw, pos= 0.833] {\(B = \ker f \cap \ker d\)} (c1 -|
                c2);
\end{tikzpicture}\nonumber
\end{equation}
\caption{Decomposition Theorem diagram.}\label{fig:decomposition}
\end{figure}

These properties are illustrated in the diagram represented in Figure \ref{fig:decomposition}.

It is a simple exercise of elementary linear algebra to prove the equivalence
between the above diagram and the initial reduction.


The Hexagonal Lemma~\cite{RS12} allows us to reduce only a module of
a chain complex in a particular degree. It is possible to generalize
this lemma applying the reduction to every degree simultaneously.

\begin{Thm}[Generalisation of the Hexagonal Lemma]\label{thm:hex2}
Let $C_\ast=(C_n,d_n)_{n \in \Zset}$ be a chain complex. We assume
that every module is decomposed $C_n = A_n \oplus B_n \oplus C'_n$.
The differential map $d_n$ are then decomposed in $3 \times 3$ block
matrices $[d_{n,i,j}]_{1\leq i,j \leq 3}$. If every component
$d_{n,2,1}: A_n \rightarrow B_{n-1}$ is an isomorphism, then the
chain complex can be canonically reduced to a chain complex
$(C'_n,d'_n)$. The components of the desired reduction are:

\begin{equation}
\begin{array}{c}
d'_n = d_{n,3,3} - d_{n,3,1}d^{-1}_{n,2,1}d_{n,2,3}
 \;\;\;\;\;
f_n= [\begin{array}{ccc}
0 & -d_{n,3,1}d^{-1}_{n,2,1} & 1
\end{array}]\\ \\
g_n = \begin{bmatrix}{-d^{-1}_{n,2,1}d_{n,2,3}}\\{0}\\{1}\end{bmatrix} \;\;\;\;
h_{n-1}=\begin{bmatrix}{0}&{d^{-1}_{n,2,1}} & {0}\\{0}&{0}&{0} \\{0}&{0}&{0}\end{bmatrix}
\end{array}
\label{eq:xdef}\nonumber
\end{equation}
\end{Thm}

Again, it is not difficult to check that the displayed formulas satisfy the relations of a reduction stated in Definition \ref{defn:reduction}.

With these two auxiliary lemmas, it is possible to obtain a proof of
the BPL. The process is the following one. Given a reduction
$\rho=(f,g,h): (C,d) \rrdc (\wC,\widehat{d})$ and a perturbation
$\delta$ of the differential $d$ such that $\delta \circ h$ is
locally nilpotent, it is necessary to build a new reduction
$\rho'=(f',g',h'): (C, d +\delta) \rrdc (\wC, \widehat{d} +
\widehat{\delta}) $.

The reduction $\rho: C_\ast \rrdc \wC_\ast$ allows us to apply the
Decomposition Theorem and to obtain the diagram of Figure
\ref{fig:decomposition} where $C = A \oplus B \oplus C'$. Then, the
differential $(d+\delta)$ can be depicted in nine blocks following
that decomposition. If the component $(d+\delta)_{21}: A \rightarrow
B$ is invertible then the Generalisation of Hexagonal Lemma can be
applied and the BPL is proved.

We have that $(d+\delta)_{21} = d_{21} + \delta_{21} = d_{21} +
\delta_{21}\circ h_{12}\circ d_{21} = (id + \delta_{21}\circ
h_{12})\circ d_{21}$. Then, as $d_{21}$ is invertible (in fact, $h_{12}$ is its inverse),
we focus on proving that the other member of the product, namely $(id +
\delta_{21}\circ h_{12})$, is invertible.

As $\delta\circ h$ is locally nilpotent, i.e. for every $x \in C$,
there exists $m \in \mathbb{N}$ satisfying $(\delta\circ h)^m(x)=0$,
we obtain that in particular  $(\delta_{21}\circ h_{12})^m(x) = 0$
since the unique non-null component of $h$ is $h_{12}$. Then, the
inverse of $(id + \delta_{21}\circ h_{12})$ is $\sum_{i=1}^\infty
(-1)^i(\delta_{21}\circ h_{12})^i$.

\subsection{Formalisation of the BPL}

The formalisation of the proof of the BPL in SSReflect
requires to restrict the data structures to finitely generated chain
complex. These structures are presented in Subsection
\ref{subsec:struc}. Then, the Decomposition Lemma and the
Generalized Hexagonal Lemma are formalised in Subsection
\ref{subsec:decompTheorem} and \ref{subsec:genHL}. These two results are the key
ingredients used in the formalisation of the BPL -- included in
Subsection \ref{subsec:proofBPL}. Before that, we start providing a
brief explanation about the formalisation of the kernel of a map in
SSReflect. The representation chosen for this well-known
notion in mathematics has important consequences in the rest of
structures used in the formalisation of the proof.

\subsubsection{The kernel of a map}
 \label{subsc:kernel}
The kernel of a finite map is defined by the kernel of the matrix
which represents this map. This is defined in SSReflect in
the following way.

\begin{lstlisting}
  Definition kermx m n (A: 'M_(m,n)): 'M_m
    := copid_mx (\rank A) *m invmx (col_ebase A).
\end{lstlisting}

The kernel of a matrix \lstinline?A? is defined as the inverse of \lstinline{col_ebase}
(the extended column basis of \lstinline?A?), with the top \lstinline?\rank A? rows zeroed out
 -- this is achieved thanks to the multiplication of such inverse by a square diagonal matrix 
 with 1's on all but the first \lstinline?\rank A? coefficients on its main diagonal (\lstinline?copid_mx (\rank A)?).
 In other words, the kernel of a matrix $A$ is a square matrix whose row space consists of all
 \lstinline?u? such that \lstinline?u *m A = 0?. This property is expressed in the following lemma.

\begin{lstlisting}
  Lemma mulmx_ker m n (A : 'M_(m, n)) : kermx A *m A = 0.
\end{lstlisting}

Two comments are required about this definition. Firstly, the kernel consists of the elements that are made null when they are applied to the left.
In our previous mathematical definitions they are null when applied to the right. Secondly, in our development, we have chosen to delete the top
\lstinline?\rank A? null rows of the kernel. If we worked with the original definition, we would obtain partial identities instead of identities 
in some proofs. 

\begin{lstlisting}
  Definition ker_min  (m n : nat) (M : 'M_(m,n)) :=
    (castmx ((Logic.eq_refl (m-\rank M)),(\rank M) + (m-(\rank M))) = m)
    (row_mx (@const_mx _ (m-\rank M) (\rank M) 0) 1%:M)) *m (kermx M).
\end{lstlisting}

The previous definition consists of the product of a row matrix with
the kernel. The row matrix is composed by a block of zeros with
\lstinline?m-\rank M? rows and \lstinline?\rank M? columns and an
identity matrix of dimension \lstinline?\rank M?.  Let us note that
the cast (a \emph{coercion} which allows us to change an entity of 
one data type into another) in the definition is necessary so that 
the product can be properly defined -- \lstinline?(kermx M)? 
and \lstinline?(row_mx (@const_mx _ (m-\rank M) (\rank M) 0) 1%:M))?
have, respectively, types \lstinline?'M_m? and 
\lstinline?'M_(m-\rank M, \rank M + m- \rank M)?; so, we cast the
type of \lstinline?(row_mx (@const_mx _ (m-\rank M) (\rank M) 0) 1%:M))?
to \lstinline?'M_(m-\rank M, m)? using the \lstinline?castmx? function.
Anyway, both definitions generate the same space as we can see in
the following lemma.

\begin{lstlisting}
  Lemma ker_min_kermx (m n : nat) (M : 'M_(m,n)) :
    (kermx M :=: (ker_min M))%MS.
\end{lstlisting}

 \subsubsection{Main mathematical structures}
\label{subsec:struc}

Let us define a finitely generated chain complex with elements in a field.

\begin{lstlisting}
  Variable K: fieldType.
  Record FGChain_Complex :=
   {m: Z -> nat;
    diff: forall i: Z, 'M[K]_(m (i + 1), m i);
    boundary: forall i: Z, (diff (i + 1)) *m (diff i) = 0}.
\end{lstlisting}

Some comments about this definitions are necessary. The chain
complex definition contains a function denoted by \lstinline?m?
which obtains the number of generators for each degree. Then, we can
define the differentials using the matrix representation of these
maps \lstinline?forall i: Z, 'M[K]_(m (i + 1), m i)?. Two important
design decisions have been included in this definition. Due to the definition
of the kernel of a matrix in 
SSReflect we will work with transposed matrices.  This implies that
the product is also reversed. Furthermore, the degrees of the
differentials have been increased in one unit. This is an
alternative definition to the usual one which considers the
differential in degree \lstinline?i? as a function from degree
\lstinline?i+1? to degree \lstinline?i?~\cite{DR2011}. It is clear
that as we are considering the definition for all the integers, both
definitions are equivalent. However, with our version,  a Coq
technical problem is easily avoided.

Now, we can define the notions of chain complex morphism and homotopy operator
for the \lstinline?FGChain_Complex? structure.

\begin{lstlisting}
  Record FGChain_Complex_Morphism (A B: FGChain_Complex) :=
  {FG : forall i: Z, 'M[K]_(m A i, m B i);
   FG_well_defined: forall i: Z, (diff A i) *m (FG i) = (FG (i+1)) *m (diff B i)}.
   
  Record FGHomotopy_operator (A: FGChain_Complex) :=
  {Ho: forall i:Z, 'M[K]_(m A i, m A (i+1)%Z)}.
\end{lstlisting}

With these previous structures, we can define the notion of a
reduction for a finitely generated chain complex. Namely, it is a record
\lstinline?FGReduction? with two chain complexes
\lstinline?C D: FGChain_Complex?, two morphisms
\lstinline?F: FGChain_Complex_Morphism C D?,
\lstinline?G: FGChain_Complex_Morphism D C?, a homotopy operator
\lstinline?H: FGHomotopy_operator C?, and the five properties of the reduction.
For instance, the second property is defined as
\lstinline?ax2: forall i: Z,  (FG F (i+1)) *m (FG G (i+1)) + (Ho H (i+1)) *m (diff C (i+1)) + (diff C i) *m (Ho H i) = 1%:M?.


 We are going to focus our
attention on the \lstinline?(diff C i) *m (Ho H i)? component of the
definition of a reduction. This product is possible without casts.
If we consider the mathematically equivalent definition with the
differential in degree \lstinline?i? as a matrix
\lstinline?'M_(m i,m(i-1))?, then the corresponding component would be
\lstinline?(diff C i) *m (Ho H (i-1))?. In this case a cast would be required to
transform elements in degree \lstinline?(i-1+1)? into elements in
degree \lstinline?i?. Using such a definition this kind of casts would populate the
development; however, using our alternative definition, we avoid the use of casts.

\subsubsection{Formalisation of the Decomposition Theorem}
\label{subsec:decompTheorem} In this subsection, we focus on proving
the Decomposition Theorem. The powerful
SSReflect library on matrices makes this development easier. Given a reduction \lstinline?rho: FGReduction K?
we define the decomposition of \lstinline?(C rho)? through
an isomorphism between the module with \lstinline?(m (C rho) i)? generators and the module built from the sum of three set of generators. The first morphism of the
isomorphism reflects a change of basis between both modules (where we work with the canonical basis in the first one) taking into account how the second one is divided:
\begin{lstlisting}
  Definition Fi_isom (i: Z):=
   (col_mx (ker_min (row_mx (FG (F rho) (i+1)) (Ho (H rho) (i+1))))
           (col_mx (ker_min (row_mx (FG (F rho) (i+1)) (diff (C rho) i)))
             (FG (G rho) (i+1)))).
\end{lstlisting}
In this definition it is necessary to note that the row space of the column matrix \lstinline?col_mx A B? is the sum of the row spaces of the matrices \lstinline?A? and \lstinline?B?, and that the intersection of kernels of two matrices \lstinline?A? and \lstinline?B? generates the same space that the kernel of a row matrix \lstinline?row_mx A B?.

The definition of the inverse morphism  \lstinline?Gi_isom? is
obtained knowing that it is a row matrix composed by three blocks
and using the second property of the reduction definition.

Then, applying the change of basis to the source chain complex of the reduction \lstinline?(C rho)? we obtain a new reduction between the decomposed chain complex and the second chain complex of the reduction \lstinline?(D rho)?.
It is not difficult to prove that the components of that reduction are:
$$\text{\lstinline?D_rho_base i?} = \left(\begin{array}{ccc}
 0 &  \text{\lstinline?D_aux i?} &0 \\
 0& 0 & 0 \\
 0 & 0 &  \text{\lstinline?diff (D rho) (i+1)?}
 \end{array} \right) \;\;\;\;\; \text{\lstinline?F_rho_base i?} = \left(\begin{array}{l}
 0 \\
 0\\
 \text{\lstinline?id?}
 \end{array} \right)$$

 $$\text{\lstinline?G_rho_base i?} = \left(\begin{array}{ccc}
 0 & 0&\text{\lstinline?id?}
 \end{array} \right)\;\;\;\;\; \text{\lstinline?H_rho_base i?} = \left(\begin{array}{ccc}
 0 & 0&0 \\
 \text{\lstinline?H_aux i?} & 0 & 0 \\
 0 & 0 &0
 \end{array} \right)$$

These components directly reflect the structure included in Figure \ref{fig:decomposition}. Finally, the two isomorphisms included in that diagram are easily obtained from these components using the properties of the redefined reduction.




\subsubsection{Formalisation of the Generalisation of the Hexagonal Lemma}
\label{subsec:genHL}

The first step in the formalisation of the Generalisation of the Hexagonal Lemma is defining its hypotheses. As every module is divided into three parts, every differential consists of nine blocks.
\begin{lstlisting}
  Variables (m1 m2 m3: Z -> nat).
  Variable Di: forall i: Z,
    'M[K]_(m1(i+1)+(m2(i+1)+m3(i+1)), (m1 i)+((m2 i)+(m3 i))).
  Hypothesis boundary: forall i:Z, Di (i+1) *m Di i = 0.
  Definition CH:= Build_FGChain_Complex boundary.
\end{lstlisting}
 We recall that we are working with transposed matrices. Consequently, the block $(1,2)$ of each differential, denoted by \lstinline?d12?, will be an isomorphism.
\begin{lstlisting}
  Variable d12_1: forall i: Z, 'M[K]_(m2 i, m1 (i + 1)).
  Hypothesis d12_invertible:
    forall i: Z, (d12 (i+1)) *m (d12_1 (i+1)) = 1%:M /\
                (d12_1 (i+1)) *m (d12 (i+1)) = 1%:M.
\end{lstlisting}

Afterwards, we define the morphisms \lstinline?fi? and
\lstinline?gi? and the homotopy operator \lstinline?hi? to build a
reduction of the chain complex \lstinline?CH?. These maps are
detailed in the proof of Theorem \ref{thm:hex2}. For instance, we
define \lstinline?gi i = ((-(d32 i)*(d12_1 i))  0 1)?. Let us remark
that \lstinline?(gi i)? is a matrix whose type is:
\lstinline?'M_(m3(i+1), m1(i+1)+m2(i+1)+m3(i+1))?. In this way
\lstinline?(gi i)? is defined between the modules of degree
\lstinline?(i+1)? instead of the ones of degree \lstinline?i?. This
allows us to avoid casts as in the case of the definition of the
differential. Then, the rest of the components of the reduction are
defined accordingly. Finally, the proof of the reduction properties
are obtained using essentially rewriting tactics, allowing us to
build the required reduction \lstinline?rhoHL?.

\subsubsection{Putting together the pieces to obtain a formalisation of the BPL}
\label{subsec:proofBPL}

In order to obtain a formalisation of the BPL, we, firstly, define the hypotheses of the lemma. Those are a reduction and a perturbation of the first chain complex of the reduction.
\begin{lstlisting}
  Variable K: fieldType.
  Variable rho: FGReduction K.
  Variable delta:  forall i:Z, 'M[K]_(m (C rho)(i+1), m (C rho) i).
  Hypothesis boundary_dp: forall i:Z,
    (diff (C rho) (i+1) + delta (i+1)) *m (diff (C rho) i + delta i) = 0.
\end{lstlisting}

In addition, we will assume the nilpotency hypothesis. Let us note
that \lstinline?pot_matrix? is a function which we have defined to
compute the power of a matrix.
\begin{lstlisting}
  Variable (n: nat).
  Hypothesis nilpotency_hp:  forall i: Z,
    (pot_matrix (delta i *m Ho (H rho) i) n = 0).
\end{lstlisting}

With these hypotheses, we have to define a reduction from the chain
complex with the differential of \lstinline?(C rho)? perturbed by
\lstinline?delta?. Firstly, we apply the Decomposition Lemma over
the reduction \lstinline?rho?, this allows us to decompose each
\lstinline?(diff (C rho) i)? in the $9$ blocks given in the proof of
the lemma (see Subsection \ref{subsec:decompTheorem}). The
isomorphism given by \lstinline?Fi_isom? and \lstinline?Gi_isom? are used 
to decompose the perturbation in $9$ blocks.

\begin{lstlisting}
  Definition deltai_new (i:Z):=
    (Fi_isom rho (i+1)) *m (delta (i+1)) *m (Gi_isom rho i).
\end{lstlisting}


Let us note that the differential \lstinline?deltai_new? has moved
up one degree with respect to \lstinline?delta?. In this way, we
obtain a division in 9 blocks of the chain complex
\lstinline?Di_pert i:= diff (C rho) (i+1) + delta(i+1)?.




Then, the Generalisation of the Hexagonal Lemma can be applied if the block $(1,2)$
of that chain complex is invertible.
Following the chain of equalities included in the mathematical proof of this lemma,
it is enough to prove that \lstinline?1%:M + H_aux * deltai_new21?
has an inverse. To this aim, the following lemma is useful -- the lemma is proved using the
powerful \lstinline?bigop? library of SSReflect~\cite{BGOP08}.

\begin{lstlisting}
  Lemma inverse_I_minus_M_big (M: 'M[R]_n): (pot_matrix M m = 0) ->
    (\sum_(0<=i<m) (pot_matrix M i)) *m (1%:M - M) = 1%:M.
\end{lstlisting}

Now, applying the Generalisation of the Hexagonal Lemma
\lstinline?rhoHL? we build a reduction \lstinline?quasi_bpl? of
the decomposed and perturbed chain complex.

The Generalisation of the Hexagonal Lemma proved in Subsection
\ref{subsec:decompTheorem} builds a reduction of the chain complex
given as hypothesis but moved up in a degree. Moreover, the
definition of \lstinline?Di_pert? has required us to define it moved
up in a degree more. To sum up, we have built a reduction from a
perturbed chain complex but moved up two degrees
\lstinline?Di_BPL_up i:= (C rho)(i+1+1)+ delta(i+1+1)?. Finally, in
order to obtain a reduction \lstinline?rho_BPL? from the initial
chain complex \lstinline?(C rho)? perturbed by \lstinline?delta?, we
can move down two degrees in the reduction obtained. For instance,
the differentials are defined as
\lstinline?Di_BPL i:= castmx(cast3 i,cast4 i)(Di_BPL_up(i-1-1))?. In this way, we have avoided the use of
casts derived from the degrees until the last step of the proof.

\subsection{Using the BPL to reduce the chain complex associated with a digital image}
\label{sec:vf_bpl}

Different ways to represent matrices exist in a system. In SSReflect, there are two
available representations. The first one formalises a matrix as a
function. This function determines every element of the matrix
through two indices (for its row and its column). With this abstract
representation it is not difficult to define different operations
with matrices and prove properties of them. Indeed, an extensive
library on matrices is provided in SSReflect. For this reason,
this representation was chosen to prove the BPL. However, this
representation is not directly executable, since this matrix definition is locked to avoid the trigger of heavy
computations during deduction steps. To overcome this drawback, an alternative 
definition which represents matrices as lists of lists was introduced in~\cite{Ex_rank}. 
This concrete representation allows us to define operations which can be executed within Coq.
Due to this reason, this was the representation chosen for the implementation of the RS algorithm, see Subsection~\ref{subsec:frss}.
In the negative side, proving properties using this representation is
much harder, and we do not dispose of the extensive SSReflect library.


In order to combine the best of both representations, a technique was introduced in~\cite{Ex_rank}. 
In that work, two morphisms are defined: \lstinline?seqmx_of_mx? from
abstract to concrete matrices and \lstinline?mx_of_seqmx? from
concrete to abstract matrices. The compositions of this morphisms are
identities. In this way, it is proved that these two matrix
representations are equivalent. Besides, these morphisms allow
changing the representation when required. We are going to use that
idea to prove the Vector Field Reduction Theorem for a chain complex
generated from a digital monochromatic image.

The process begins by defining a new structure to represent a
particular type of chain complexes, called by us \emph{truncated
chain complex}. They consist only of two matrices
\lstinline?d1, d2: matZ2?, whose product is null.
These matrices are the only not null components in the differential
map of a chain complex. In this way, the null elements of this chain complex are not
included in this definition. Besides, these matrices are represented using computable
structures. The companion notions of truncated chain complex morphism,
truncated homotopy operator, and truncated reduction are also defined for this structure.




As an example of the previously mentioned technique, we include the
following definitions of two functions to sort a matrix,
one for matrices represented as lists, and another one for matrices
represented as functions (an equivalence between them is also provided
in our development).

\begin{lstlisting}
  Definition reorderM (s1 s2: seq nat)(M: matZ2):=
   (take_columns_s s2 (take_rows_s s1 M)).
\end{lstlisting}

\begin{lstlisting}
  Definition reorder_mx (s1:'S_m)(s2:'S_n)(M:'M[R]_(m,n)):=
    col_perm s2 (row_perm s1 M).
\end{lstlisting}

Although these two definitions seem similar they adopt different
approaches. The first one uses simple types which are closer to a
standard implementation in traditional programming languages.
Indeed, that implementation is a direct translation of the one
made in Haskell.  We will use this version to reorder the
structures when we need to compute with them. For instance, let
\lstinline?chaincomplexd1d2? be an
initial truncated chain complex which
differential is given by \lstinline?d1: matZ2?, with \lstinline?m?
rows and \lstinline?n? columns, and \lstinline?d2: matZ2? with
\lstinline?n? rows and \lstinline?p? columns. It corresponds with
the representation of the initial digital image that we want to
reduce. The first definition is used to obtain the ordered list of
lists \lstinline?d1'? and \lstinline?d2'? after computing an ordered
and admissible discrete vector field for \lstinline?d1?.

The second definition uses the full power of dependent types and
the structures and properties developed in the SSReflect library.
We will use this version
when we need to prove properties on the reordered structures. For
instance, this definition is used to obtain the boundary property of
the truncated chain complex \lstinline?chaincomplexd1'd2'?, or to
define an isomorphism between \lstinline?chaincomplexd1d2? and
\lstinline?chaincomplexd1'd2'?. Both proofs take profit of
properties on permutations included in the library.

We introduce just another example of the use of this approach. We
need to prove that the upper-left submatrix of \lstinline?d1'? is a
lower triangular matrix (of dimension the number of vectors in the
admissible vector field, \lstinline?m1?) with 1's on its diagonal.
In this case, the proof needs reasoning on the functions included in
the RS algorithm and a quite long battery of add-hoc lemmas on the
computable structures are required. In a second step, we need to
prove that this matrix has an inverse. This second proof is easy using the results included in the SSReflect library
after changing the representation.

Now, if we want to apply the BPL, we need to build a chain complex from
the ordered truncated chain complex \lstinline?chaincomplexd1'd2'?.
This process requires some technical steps:  translate lists to 
SSReflect matrices, transpose the matrices, and complete with null
matrices all the not provided components. In particular, from
\lstinline?d'1? we define the following matrix \lstinline?d'1_trmx?
which type is \lstinline?'M[K]_(m1+(m - m1),m1+(n - m1))?:
\begin{lstlisting}
  Definition d'1_trmx :=
    trmx((mx_of_seqmx (m1 + (m - m1)) (m1 + (n - m1)) d'1)).
\end{lstlisting}

Then, we build the differential of a chain complex, denoted by
\lstinline?CC_ordered?, in the following way:
\begin{lstlisting}
  Definition diff_m:= fun i:Z => match i as
   z return ((fun z0 : Z => 'M_(m_m (z0 + 1), m_m z0)) z) with
    |(-1)%Z => d0_m
    |0%Z => d'1_trmx
    |1%Z => d'2_trmx
    |2%Z => d3_m
    |3%Z => dn_m
    |_ => dn_m
  end.
\end{lstlisting}
We have to highlight the differentials \lstinline?d0_m? and
\lstinline?d3_m? because they are matrices with rows and no columns or
with columns and no rows. The rest of differentials will be matrices
with no rows and no columns.

We will obtain a reduction of this initial chain complex
\lstinline?CC_ordered? applying the BPL to the following auxiliary
reduction. Let us define a chain complex \lstinline?Dcc? whose
modules have the same rank than the modules of
\lstinline?CC_ordered?,  and whose differential has only one not
null component.  This component of type
\lstinline?'M[K]_(m1+(m-m1),m1+(n-m1))? is defined by a matrix \lstinline?hat_d1? whose
upper-left block is the identity matrix of dimension \lstinline?m1?
and is null in the rest of its blocks. The reduced chain complex
\lstinline?Ccc? has modules whose ranks are determined by the
critical cells found by the RS algorithm on \lstinline?d1?. That is,
the only not zero ranks are \lstinline?m-m1?, \lstinline?n-m1?,
and \lstinline?l?. The differential of this reduced chain complex is
null.  Then,  it is easy to build a
reduction between \lstinline?Dcc? and \lstinline?Ccc? having
\lstinline?h_0? (extracted from the homotopy operator given by
\lstinline?trmx hat_d1?) as its unique non-null component.


Now, we define a perturbation \lstinline?delta_m?  of \lstinline?Dcc?. The non-null components of this perturbation are:
\begin{lstlisting}
  Definition delta_1 := d'1_trmx - hat_d1.
  Definition delta_2 := d'2_trmx.
\end{lstlisting}


Then, if we prove the nilpotency condition we can apply the BPL.
This lemma directly obtains the required reduction of
\lstinline?CC_ordered?. The natural number which allows us to prove
this property is \lstinline?m1.+2?. As we are working with finite
structures, this number can be taken as a uniform bound for which the
nilpotency condition is verified for all the elements in the module.
\begin{lstlisting}
  Lemma nilpotency_hp_m : forall i:Z,
    (pot_matrix(delta_m i *m (Ho (H reductionFG_gen) i))(m1.+2) = 0).
\end{lstlisting}

In the proof of this lemma, the only interesting case is when
\lstinline?i=0?, that is:
\lstinline?pot_matrix (delta_1 *m h_0)(m1.+2) = 0?. Expanding the definitions using blocks, we obtain that
the only non-null blocks are
\lstinline?pot_matrix (d11 - id)(m1.+2)? and \lstinline? d21 *m pot_matrix(d11 - id) (m1.+1)?, where
\lstinline? d11?  and \lstinline? d21? are, respectively, the up-left and
down-left blocks of \lstinline?d'1_trmx?.

Then, it suffices to prove that
\lstinline?pot_matrix(d11 - id)(m1.+1)=0?. Let us recall that \lstinline?d11? is an upper
triangular matrix with type \lstinline?'M_m1?. We define the
notion \lstinline?upper_triangular_up_to_k? which given a natural
number \lstinline?k? and a matrix \lstinline?M? determines whether the
matrix is an upper triangular matrix with 0's under and on the
\lstinline?k?-th diagonal. This definition is formalised in 
SSReflect as follows.
\begin{lstlisting}
  Definition upper_triangular_up_to_k n k (M: 'M[F]_n) :=
    forall (i j: 'I_n), j <= i + k -> M i j = 0.
\end{lstlisting}

We have developed a small library of properties on upper triangular
matrices which includes the following generalisation of the lemma
to prove.
\begin{lstlisting}
  Lemma upper_triangular_pot_matrix_n n (M: 'M[F]_n) :
    upper_triangular_up_to_k 0 M -> (pot_matrix M n.+1) = 0.
\end{lstlisting}

The BPL allows us to define a reduction \lstinline?red_BPL? of
\lstinline?CC_ordered?. In order to obtain a truncated
reduction from this reduction we retrace the technical steps made to
apply the BPL: extract from this reduction the non-null maps,
apply the transpose operation, and change the matrix
representation.  For instance, the two components of the first
truncated chain complex in the reduction are defined in the following way.
\begin{lstlisting}
  Definition d'1_trmx_trmx:= seqmx_of_mx (trmx (diff (C red_BPL) 0)).
  Definition d'2_trmx_trmx:= seqmx_of_mx (trmx (diff (C red_BPL) 1)).
\end{lstlisting}
It is necessary to stress that the obtained big truncated chain complex is composed of the differentials where the transpose operation has been applied twice. Some casts are needed in a last step in order to build a truncated reduction of the chain complex \lstinline?chaincomplexd'1d'2?.

Finally, to obtain the required reduction, it remains to compose the isomorphism between the truncated
chain complex \lstinline?chaincomplexd1d2? obtained from a digital
image to the ordered truncated chain complex
\lstinline?chaincomplexd1'd2'? with the reduction built as explained in the last
paragraph.

\section{Certified computation of the homology associated with a digital image}\label{computation}

We have previously explained that we cannot directly compute in
Coq the homology of a chain complex produced by actual images
as the one of Figure~\ref{fig:neuron} due to its size. Nevertheless, this
computation is possible with the reduced chain complex. From a
theoretical point of view, our objective consists in developing a
formal proof which verifies that both images have the same homology;
this has been accomplished in the previous sections. But, as we are working in a
constructive setting, we could also undertake the computation  
inside Coq of the reduced chain complex. In the development included 
in the previous sections, only the first step towards that reduction is
executable: the calculation of an admissible discrete vector field. 
The reduced matrix is then obtained through the BPL, which uses not 
executable matrices (because we have worked with the matrix definition 
included in the SSReflect library in order to prove the BPL).

We have developed a proof of the BPL for general (finitely generated) chain complexes,
providing a notable formalisation. Then we have applied it in a very
particular situation: we consider truncated chain complexes obtained
from digital images, and we obtain an admissible discrete vector field
using the RS algorithm on one of the matrices $d1$ of the
differential map. In this situation, it is possible to obtain a
reduction of $d1$ using directly the Hexagonal Lemma~\cite{RS12}.
This reduction is extended to the other matrix of the differential
in order to preserve the boundary condition. Concretely, the initial
matrix $d1$ is divided into 4 blocks and the reduced one is built
by the formula $d22 - d21*d11^{-1}*d12$. In order to obtain an
executable version of that reduction we have also formalised this
simplified proof using computable structures. With this version we
are able to compute the whole process for toy examples. 

When the size of the matrix grows, it is not possible to compute the full path directly inside Coq. 
But we can calculate, in a certified manner within Coq, the homology of the reduced matrix. 
That is the case of actual images as the one in Figure~\ref{fig:neuron}. In this case, 
the original matrix has dimensions $690\times 1400$ and the reduced one $97 \times 500$. 
This reduced matrix has been obtained using the Haskell version of the algorithm formalised in Coq.
With the current state of the running machinery in Coq, the system is not able to compute of the steps producing 
this reduced chain complex. The bottleneck
of our development is the definition of the inverse of a matrix.
Although we have used a specialized function to compute the inverse of a
triangular matrix with 1's on its diagonal, implemented in \cite{Seqmatrix}, the performance gain has not been enough in this particular application.
Finally, the solution adopted was to use Haskell as an oracle.
The process is the following. From the matrices of the chain complex of a digital image, we delegate in Haskell the computation of
the reduced matrix and the matrices of the required reduction (they also include the inverse matrix in their definition).
Then, these matrices are brought into Coq which obtains the homology (in 25 seconds) of the reduced matrix and automatically
(using the \lstinline?vm_compute? tactic) compute the proofs (in 8 hours) of the reduction between both matrices.


\section{Conclusions and further work}

In this paper, we have reported on a research providing a certified
computation of homology groups associated with some digital images
coming from a biomedical problem. The main contributions allowing
us to reach this challenging goal have been:

\begin{itemize}

\item The implementation in Coq of Romero-Sergeraert's algorithm
\cite{RS12} computing a discrete vector field for a digital image.

\item The complete formalisation in Coq/SSReflect of a proof of
the Basic Perturbation Lemma (to be compared with the one 
at \cite{ABR08}).

\item A methodology to overcome the problems of efficiency related
to the execution of programs inside proof assistants; in this approach
the programming language Haskell has been used with two different
purposes: as a fast prototyping tool and as an oracle.

\end{itemize}

As for future work, several problems remain open. The most evident
one is obtaining a better performance in the process. This can be undertaken 
at three different levels. First, by using other algorithms to compute the main
objects in our approach (discrete vector fields of digital images, inverses
of matrices, and so on). Second, by implementing more efficient data structures,
suitable for the theorem proving setting. Third, by improving the running
environments in proof assistants.

Another line of research is to apply our methodology and techniques to
other problems related to the homological processing of biomedical images.
The best candidate is \emph{persistent homology}, which has been already
applied and formalised (see \cite{TCL}). The project would be to study whether
our reduction strategy can be also profitable in this new homological context. 

\section*{Acknowledgements}
 Partially
 supported by Ministerio de Educaci\'on y Ciencia, project
 MTM 2009-13842-C02-01, and by the FORMATH project, nr. 243847,
 of the FET program within the 7th Framework program of the European Commission.

\bibliographystyle{plain}
\bibliography{crship}

\end{document}